\documentclass[aps,prmaterials,reprint,superscriptaddress,floatfix]{revtex4-2}

\usepackage{amsmath}
\usepackage{amssymb}
\usepackage{bm}
\usepackage{graphicx}
\usepackage{float} 
\usepackage{textcomp}
\usepackage{gensymb}
\usepackage[colorlinks=true, allcolors=blue]{hyperref}
\DeclareFontFamily{U}{wncy}{}
\DeclareFontShape{U}{wncy}{m}{n}{<->wncyr10}{}
\DeclareSymbolFont{mcy}{U}{wncy}{m}{n}
\DeclareMathSymbol{\Sh}{\mathord}{mcy}{"58}
\newcommand{\kbf}{\mathbf{k}} 
\newcommand{\rbf}{\mathbf{r}}
\newcommand{\abf}{\mathbf{a}}
\newcommand{\bbf}{\mathbf{b}}
\newcommand{\zhat}{$\hat{\mathbf{z}}$}
\newcommand{\kzhat}{$\hat{\mathbf{k}}_z$}
\newcommand{\Sha}{\Sh_{\abf_1,\abf_2}}
\newcommand{\Shb}{\Sh_{\bbf_1,\bbf_2}}
\newcommand{\invAA}{\AA$^{-1}$}
\begin{document}
\title{Stacking, Strain, \& Twist in 2D Materials Quantified by 3D Electron Diffraction}

\author{Suk Hyun Sung}
\author{Noah Schnitzer}
\affiliation{Department of Materials Science and Engineering, University of Michigan, Ann Arbor, Michigan 48109, USA}
\author{Lola Brown}
\affiliation{Intel Electronics, Kiryat Gat 82109, Israel}
\author{Jiwoong Park}
\affiliation{Department of Chemistry, University of Chicago, Chicago, Illinois
60637, USA}
\affiliation{Institute for Molecular Engineering, University of Chicago, Chicago, Illinois
60637, USA}
\affiliation{James Franck Institute, University of Chicago, Chicago, Illinois
60637, USA}
\author{Robert Hovden}
\affiliation{Department of Materials Science and Engineering, University of Michigan, Ann Arbor, Michigan 48109, USA}
\affiliation{Applied Physics Program, University of Michigan, Ann Arbor, Michigan 48109, USA}
\email{hovden@umich.edu}
\date{\today}

\begin{abstract} 
The field of two-dimensional (2D) materials has expanded to multilayered systems where electronic, optical, and mechanical properties change---often dramatically---with stacking order, thickness, twist, and interlayer spacing \cite{splendiani_emerging_2010,xia_spectroscopic_2015,meyer_structure_2007,kim_stacking_2013,cao_unconventional_2018}. For transition metal dichalcogenides (TMDs), bond coordination within a single van der Waals layer changes the out-of-plane symmetry that can cause metal-insulator transitions \cite{splendiani_emerging_2010,acerce_metallic_2015} or emergent quantum behavior \cite{Moore_2017}. Discerning these structural order parameters is often difficult using real-space measurements, however, we show 2D materials have distinct, conspicuous three-dimensional (3D) structure in reciprocal space described by near infinite oscillating Bragg rods. Combining electron diffraction and specimen tilt we probe Bragg rods in all three dimensions to identify multilayer structure with sub-Angstrom precision across several 2D materials---including TMDs (MoS$_2$, TaSe$_2$, TaS$_2$) and multilayer graphene. We demonstrate quantitative determination of key structural parameters such as surface roughness, inter- \& intra-layer spacings, stacking order, and interlayer twist using a rudimentary transmission electron microscope (TEM). We accurately characterize the full interlayer stacking order of multilayer graphene (1-, 2-, 6-, 12-layers) as well the intralayer structure of MoS$_2$ and extract a chalcogen-chalcogen layer spacing of 3.07 $\pm$ 0.11~\AA. Furthermore, we demonstrate quick identification of multilayer rhombohedral graphene.
\end{abstract}
\maketitle

\section{Introduction}
\begin{figure*}
    \centering
    \includegraphics[width=1\linewidth]{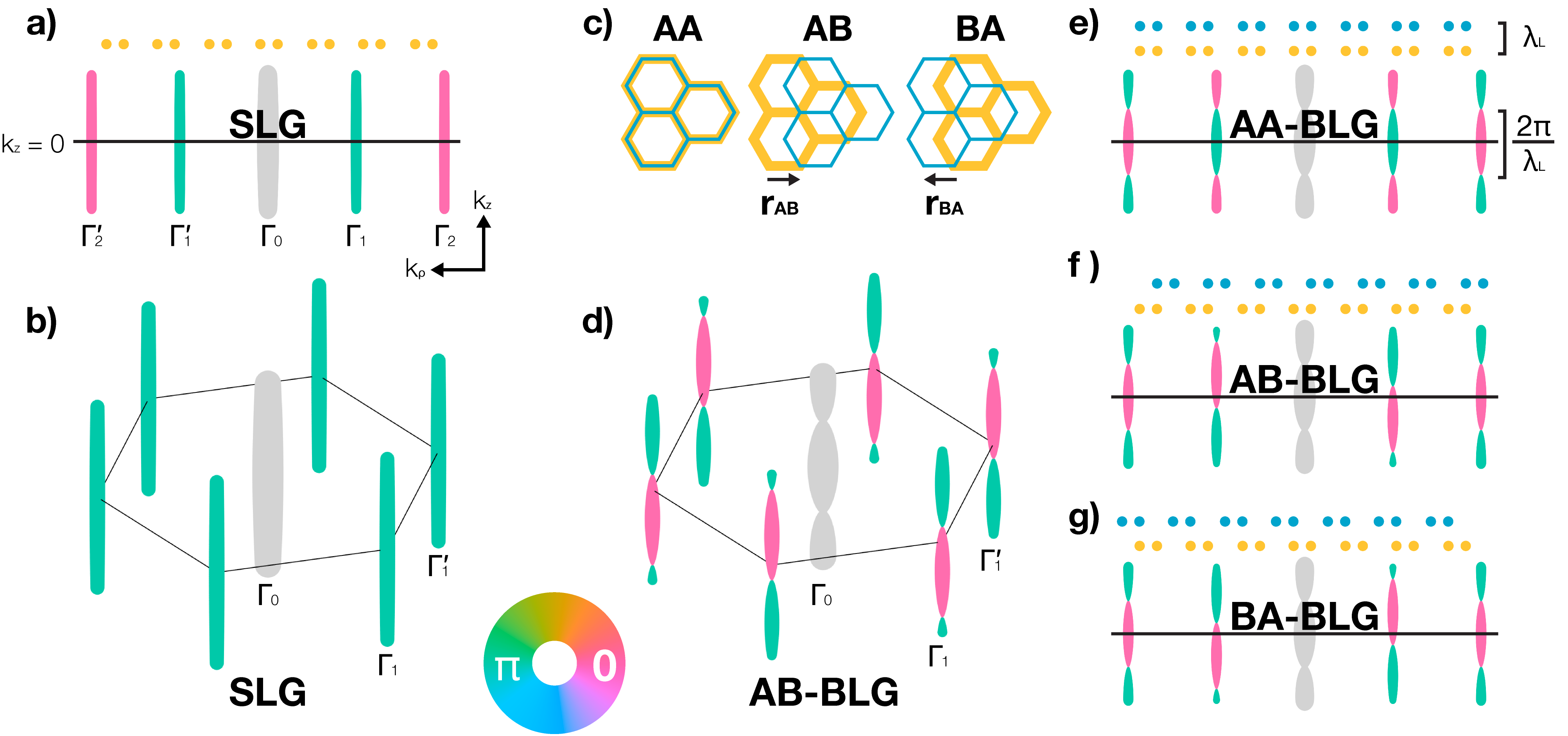}
    \caption{\textbf{3D reciprocal structure of Single and Bilayer Graphene.} 3D reciprocal space structure of a,b) single layer graphene (SLG). Width and color of Bragg rods indicate magnitude and phase (magenta = 0, teal = $\pi$), the hexagon marks $k_z = 0$ plane. c) Bilayer graphene (BLG) illustrated for AA, AB, and BA registry along \zhat. d) 3D $k$-space structure of AB-BLG. Sinusoidal magnitude---signature of multilayer systems--- is apparent. The structure of e) AA, f) AB and g) BA from side view are shown for both first ($\Gamma_1$) and second ($\Gamma_2$) order Bragg rods along with atomic stacking along $<$100$>$ direction in real-space. The barely visible decay in rod magnitude seen in SLG is due to the finite size of carbon atoms. The rod structure of BLG’s are sinusoidal with symmetry identical to the real-space. 6-fold symmetry of SLG and inversion symmetry of Bernal BLG is clearly shown in $k$-space. All structures are centered around the inversion center in real-space to maximize interpretability.}
    \label{fig:graphene}
\end{figure*}
The characteristics of layered two-dimensional (2D) materials and heterostructures are intimately linked with stacking order, as thickness and interlayer registry dramatically alter the confinement and symmetry of the system. For instance, inversion symmetric monolayer 1T-MoS$_2$ is metallic \cite{acerce_metallic_2015} while mirror symmetric monolayer 2H-MoS$_2$ is a direct band gap semiconductor \cite{splendiani_emerging_2010}. In several 2D systems, the intrinsic inversion asymmetry or symmetry breaking via external perturbation bear possibilities for electronic switching \cite{Castro2011,Ohta2006} or valleytronic devices \cite{schaibley_valleytronics_2016}. Recently, superconductivity was observed in bilayer graphene when the interlayer twist is tuned to a `magic angle' \cite{cao_unconventional_2018}.

High-precision characterization of stacking order, interlayer spacing, twist, and roughness is paramount to harnessing the diversity of 2D phenomena. The field of 2D materials erupted with facile identification of single layer graphene when exfoliated onto $\sim$300 nm thick SiO$_2$ substrates \cite{Blake_makingGrapheneVisible}. Since then, thickness characterization techniques have expanded to Raman spectroscopy \cite{ferrari2006raman}, atomic force microscopy \cite{novoselov_electric_2004}, and electron microscopy \cite{ruoff_2009}. Thickness alone, however, provides an incomplete picture of the atomic structure and stacking order. Scanning transmission electron microscopy (STEM) can image thickness with atomic resolution \cite{QuentinPRL,huang2011}, yet, this real-space projection of the specimen loses out-of-plane information, poorly discerns light elements bonded to heavy elements, and requires high radiation doses. Furthermore, a fundamental trade-off between resolution and field-of-view limits atomic resolution imaging to small regions of interest, typically (20 nm)$^2$. In contrast, electron diffraction remains a longstanding powerful tool for obtaining a representative average of the atomic structure across large areas, at lower doses, with high-throughput and high precision. 

We demonstrate electron diffraction is particularly apt for probing the three-dimensional (3D) structure of 2D materials. Contrary to the confined real-space structure, we show 2D materials have striking, measurable features in the third dimension of reciprocal space that describe key structural parameters such as stacking order, twist, strain, chemistry, and inter- or intra- layer spacing. In 2D materials, Bragg peaks extend into near infinite rods running perpendicular to the specimen surface. Each Bragg rod oscillates with intensity and phase described by the atomic arrangement within and between each 2D layer. 

Prominent distinctions arise in the reciprocal ($k$) space structure of 2D materials: 
$\circ$ In-plane lattice strain moves the position of Bragg rods.
$\circ$ Rod oscillation frequencies are inversely proportional to inter- and intra-layer spacing.
$\circ$ Out-of-plane strain changes the oscillation frequency.
$\circ$ Symmetry and structure of first order rods ($\Gamma_1$) reveal stacking order.
$\circ$ Second order ($\Gamma_2$) facilitates thickness determination.
$\circ$ TMD chemistry changes the amplitude of oscillations.
$\circ$ Twisted layers are described by a superposition of diffraction patterns for non-overlapping (incommensurate) Bragg rods.
$\circ$ Progressive broadening of rods is associated with out-of-plane micro-corrugation and stiffness.
$\circ$ Curvature of the Ewald sphere results in a small, measurable excitation error in the diffraction pattern that breaks expected Friedel symmetry. 

Combining specimen tilt and diffraction, we construct `diffraction tilt-patterns' which measure the 3D structural details of single and multilayer 2D materials. This work substantially extends previous work for few-layer graphene \cite{brown_twinning_2012,FuhrerNanoLett} and boron nitride \cite{kim_stacking_2013} to transition metal dichalcogenides (TMDs) and multilayer materials. Furthermore, we expound the foundational details required to enable a wide range 3D diffraction analysis across all 2D materials.

\section{background to diffraction of \\ 2D materials}
The wave behavior of matter was first hypothesized by de Broglie in 1924 \cite{deBroglie_1925}, and three years later validated by Thomson, Davisson, Germer with the experimental demonstration of electron diffraction \cite{thomson_diffraction_1927,davisson_reflection_1928}. In the far-field, diffracted high-energy electrons are described by a near planar slice through the specimen's 3D reciprocal structure: $V(k_z=0)$, i.e. a Fourier transform of the projected specimen potential. This kinematic approximation~\cite{Hirsch1960} accurately describes diffraction of 2D materials much thinner than the mean free path (e.g. $\ll$ 150 nm for 200 keV electron in Si~\cite{inelasticMFP}) where multiple scattering is negligible. Tilting the specimen changes the electron beam's angle of incidence, rotating the planar slice through the reciprocal lattice to probe the 3D structure. In diffraction, only squared magnitude, $|V(\kbf)|^2$, without complex phase is measured.

We are challenged to discern the third dimension of 2D materials in real and reciprocal space. Graphene is an archetypal 2D crystal where sp$^2$ bonding forms a hexagonal lattice lying within a single plane. Graphene's real-space lattice, $V_g(\rbf) = \Sha(\rbf)\delta(z)\circledast\sum_j f(\rbf-\rbf_j)$, is described by two lattice vectors, $\abf_1$, $\abf_2$, with magnitude $a_g=2.46$~\AA, and a two atom sublattice at $\rbf_j$ ($j \in 0,1$) that mimics a honeycomb. The corresponding reciprocal lattice of graphene defines Bragg rods spaced $b_g = \frac{4\pi}{a_g\sqrt{3}}$ = 2.949 \invAA apart and is described by :
\begin{align}
    V_g(\kbf) & = \Shb(\kbf)\cdot S_g(\kbf)
\end{align}

 where the complex magnitude is determined by structure factor $S_g(\kbf) = \textstyle\sum_j f(\kbf)e^{-i\kbf \cdot \rbf_j}$.
For graphene, the single atomic plane, with near infinite confinement along \zhat~ (Fig.~\ref{fig:graphene}a-\textit{top}), has a reciprocal structure with near infinite extent out-of-plane along \kzhat (Fig.~\ref{fig:graphene}a-\textit{bottom}). 
Similar elongated rel-rods arise from planar shape factors \cite{ReesShapeFunction,Cowley_ZnO} that have been studied in surface layer diffraction experiments on bulk materials and thin-films \cite{Henzler_LEED,LagallyRheed,RobinsonPRB_crystalTruncation}. Supplemental Section II discusses $\Sh(\kbf)$ formulation and normalization prefactor \cite{SuppRef}.

Therefore, 2D materials have Bragg peaks that stretch into rods. Figure \ref{fig:graphene}a,b shows single layer graphene (SLG) in reciprocal space. Its $k$-lattice has 6-fold rotational symmetry (Fig.~\ref{fig:graphene}b), reflecting the real-space symmetry at the inversion center. The rod intensity decays slowly from the origin due to the small but finite size of each atom (described by atomic scattering factor $f(\kbf)$). The attenuating magnitude reaches 80\% by 0.038 \invAA. Both first ($\Gamma_1$) and second order ($\Gamma_2$) rods are shown in Figure \ref{fig:graphene}a. For SLG, the more distant second order Bragg rods have $\sim$94\% of the squared magnitude of the first order rods.

Combining specimen tilt and diffraction allows quantification of each Bragg rod's 3D structure. In a `diffraction tilt-pattern', diffraction peaks are quantified across specimen tilt angles. As the specimen is tilted about an axis perpendicular to the beam direction, the diffraction plane rotates through the reciprocal rods of the material as shown in Figure~\ref{fig:blg}a for the first order rods of bilayer graphene. Figure~\ref{fig:blg}b illustrates the resulting tilt pattern, and the inset notes the specimen tilt axis.

Diffraction peaks both move and broaden when a 2D crystal is tilted and must be handled during quantification. Approaching higher tilts, peaks move outward from the axis of rotation---giving the illusion of unidirectional strain. The increasing distance between Bragg peaks reflects the apparent contraction in real-space when a tilted 2D crystal is viewed in projection. Thus, diffraction peaks are minimally spaced apart when the 2D crystal is perpendicular to the beam (i.e. `on-axis'). Also, the effective selected area increases as tilt increases and a geometric factor of $\cos(\theta)^{-2}$ must be incorporated to the kinematic model of diffraction of large crystals.

Bragg peaks also broaden at higher tilts due to out-of-plane rippling of the material. J.C. Meyer \textit{et al.} quantified intrinsic microscopic roughing of graphene by measuring the Bragg rod precession \cite{meyer_structure_2007}. Any micro-corrugation in a 2D sheet has local orientation changes that tilt the Bragg rods. Because selected area electron diffraction (SAED) measures an average of the crystal region, the superposition of tilted Bragg rods results in broadening along \kzhat. J.C. Meyer \textit{et al.} measured Bragg rod broadening to quantify roughness of graphene and showed that suspended single layer graphene had a surface normal that varied by $\pm5$ degrees while bilayer graphene was smoother with a $\pm1$ degree variation. Their work also highlights the importance of quantifying Bragg peaks from integrated intensities---not peak maxima.

\section{bilayer graphene} \label{sec:BLG}
\begin{figure*}
    \centering
    \includegraphics[width=1\linewidth]{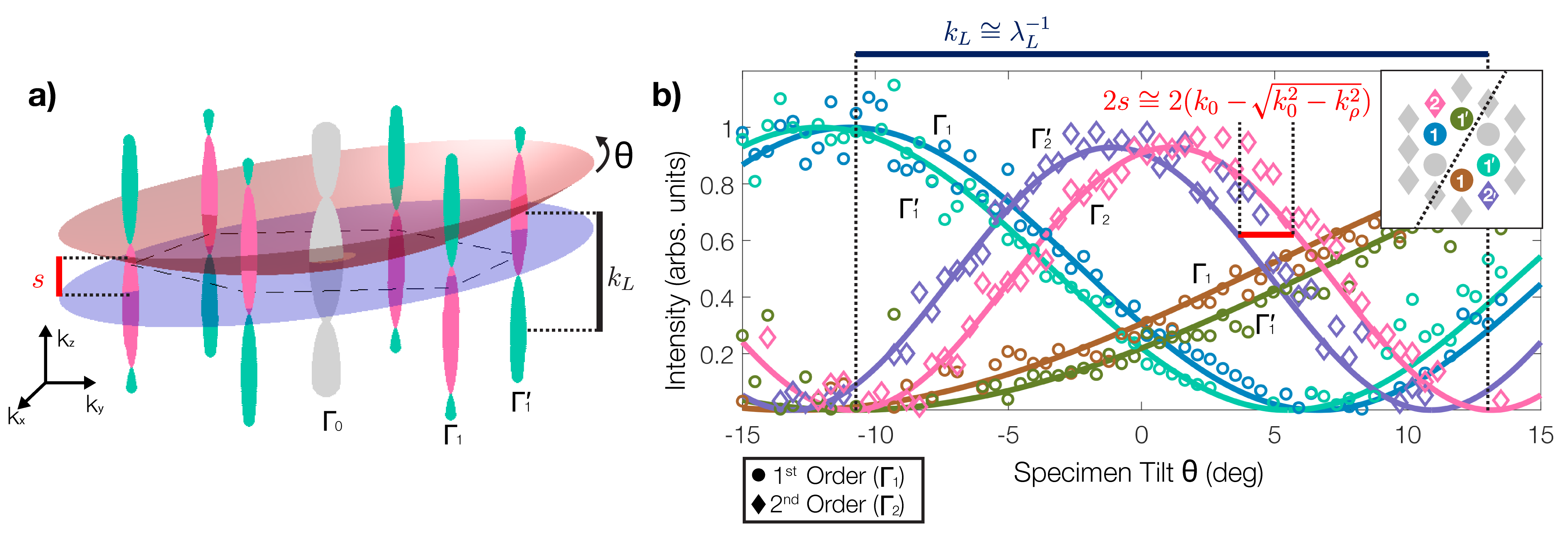}
    \caption{\textbf{Diffraction tilt-patterns of BLG.} a) 3D reciprocal rod structure of Bernal stacked bilayer graphene. The magnitude varies sinusoidally with a periodicity inversely proportional to real-space interlayer spacing ($k_{L} = \frac{4\pi}{\lambda_{L}}$). At typical TEM operation energy (blue, 200 keV), SAED is a near planar slice through the $k$-space origin; red surface exaggerates the curvature of Ewald sphere with slow electron (0.3 keV). Tilting the specimen in TEM column changes the beam's incident angle and effectively rocks the diffraction plane with respect to the Bragg rods, accessing out-of-plane information hidden in conventional TEM. The excitation error (s)---due to the curvature---is small but not negligible at low tilt angles close to the $k$-space origin. b) Kinematic ($\boldsymbol{-}$) and experimental (\textopenbullet, $\lozenge$) tilt-patterns of BLG. The tilt-patterns oscillates with frequency $k_L$. Non-trivial Ewald sphere curvature separates analogous 2nd order Friedel pair tilt-patterns (magenta($\Gamma_2$) and blue($\Gamma_2'$)) with phase difference associated with $s$.}
    \label{fig:blg}
\end{figure*}

Atomically registered bilayer materials have Bragg rods that sinusoidally oscillate in complex magnitude (Fig.~\ref{fig:graphene}d) with periodicity (${4\pi}/{\lambda_{L}}$) inversely proportional to the interlayer spacing, $\lambda_{L}$. The period of rod oscillation is independent of stacking order and depends only on interlayer spacing. Bilayer graphene (BLG) has reciprocal structure described by:
\begin{align}
    V_{bg}(\rbf) = \big[&\Sha(\mathbf{r-\Delta}/2) \delta(z-\lambda_{L}/2) \nonumber \\
    + &\Sha(\mathbf{r + \Delta}/2) \delta(z+\lambda_{L}/2) \big]
    \nonumber \\
    \circledast &\textstyle\sum_i f(\rbf-\rbf_i) \\
    V_{bg}(\kbf)=~&\Shb(\kbf)\big[2\cos{\big(\frac{\lambda_{L}}{2} k_z + \frac{\mathbf{\Delta}}{2} \cdot \kbf\big)}\big]\cdot S_g(\kbf)
\end{align}

where $\mathbf{\Delta}$ is the order parameter representing in-plane translation.

Changes to stacking order move Bragg rods up and down along \kzhat. More specifically, in-plane displacement of a layer, $\mathbf{\Delta}$, adds a phase shift $\frac{1}{2} \mathbf{\Delta \cdot k}$ to the sinusoidal intensity of each Bragg rod. There are three high-symmetry stacking configurations for BLG: energetically stable AB or BA (called Bernal or graphitic) and unstable AA (Fig.~\ref{fig:graphene}c)~\cite{Popov_2011}. The arrangement of the sinusoidal rods reflect the real-space symmetry. AA-BLG is defined by two aligned layers ($\mathbf{\Delta}=0$) with a mirror plane in-between (Fig.~\ref{fig:graphene}e). The reciprocal space structure matches the real-space 6-fold symmetry with a mirror plane at $k_z = 0$. In AB/BA-BLG, one layer is bond-length shifted with respect to the other along an in-plane bond direction ($\mathbf{\Delta}=\mathbf{\frac{a_1 + a_2}{3}}$) \cite{bernal_1924}. This translation breaks out-of-plane mirror symmetry and reduces the 6-fold symmetry to 3-fold. 

Figure~\ref{fig:graphene}e,f,g depicts the rod structure of AA, AB, and BA-BLG. The magnitude of each rod is described by its width and complex phase with color; magenta and teal represent 0 and $\pi$ respectively. Mirror symmetric AA-BLG has first order diffraction rods ($\Gamma_1$) centered about $k_z=0$ (Fig.~\ref{fig:graphene}c). For AB-BLG, the in-plane translation between atomic layers displaces $\Gamma_1$ and $\Gamma_1'$ Bragg rods in opposite out-of-plane directions ($\pm$\kzhat) with a $\pi/3$ phase shift (Fig. 1f, g).

$\Gamma_1$ rods reveal stacking order in 2D materials. For Bernal BLG the maximum intensity of odd order Bragg rods can only be measured by tilting the specimen (Fig. \ref{fig:blg}a). In the experimental tilt-pattern of AB-BLG (Fig. \ref{fig:blg}b), the non-symmetric first order Bragg rods are obvious. The blue $\Gamma_1$ curve decreases to zero intensity at 6 degrees tilt but reaches a maximum tilt at $-12$ degrees (also expected at 23 degrees). The brown $\Gamma_1$ rod on the other side of the rotation axis follows a similar opposite trend. Bragg rods more distant from the axis of rotation oscillate more rapidly in the tilt pattern. Here the axis of rotation passes through $\Gamma_2$ as shown in Figure \ref{fig:blg}b-\textit{inset}. In real-space, the maximum intensity of $\Gamma_1$ occurs when Bernal bilayer graphene is tilted so all atoms between layers lie atop one another when viewed along the beam direction. For AB and BA the patterns are mirrored and maximum intensity occurs when tilting opposite directions. Brown \textit{et al.} exploited this broken symmetry using specimen tilt to quickly distinguish AB and BA domains in bilayer graphene \cite{brown_twinning_2012}. For AA-BLG the maximum diffraction intensity trivially occurs at 0 degree tilt.

$\Gamma_2$ rods reveal the number of layers in multilayer graphene, but not stacking order. For untilted specimens ($k_z\approx0$), the intensity of the $\Gamma_2$ rods in the bilayer is four times that of monolayer graphene and will continue to scale with number of layers squared, $N^2$, as discussed in section \ref{sec:MLG}. Shown in Figure \ref{fig:graphene}e--g, the $\Gamma_2$ rods are identical and indiscernible for all three BLG stacking orders. $\Gamma_2$ rod intensity has a mirror symmetric maxima at $k_z=0$ that is clearly seen in the experimentally measured tilt-pattern (Fig.~\ref{fig:blg}b). The slight deviation of $\Gamma_2$ maxima from zero tilt is due to finite curvature of the Ewald sphere.

\section{Beam Energy \& the Ewald Sphere}
Elastic scattering preserves kinetic energy on the proverbial Ewald sphere in reciprocal space~ \cite{Ewald_1921}. At finite beam energies, the diffraction pattern is described by a spherical surface cutting through the reciprocal lattice. At typical TEM energies (60--300 keV), the curvature of the Ewald sphere is small but not negligible. As shown in Figure \ref{fig:blg}a, the Ewald sphere passes through Bragg rods slightly above the $k_z=0$ plane (historically referred to as excitation error, $s$). Tilting the specimen is equivalent to tilting the Ewald sphere. 

Diffraction tilt-patterns come in Friedel pairs~\cite{Friedel1913} comprised of a Bragg rod (at $\kbf$) and its centrosymmetric pair (at -$\kbf$). For a flat Ewald sphere (infinite beam energy) the Friedel pairs have equivalent tilt-patterns. However, with Ewald curvature the tilt patterns for each Friedel pair bifurcate with increasing separation at lower beam voltages (higher curvature). Figure \ref{fig:blg}b shows the measurable curvature of the Ewald sphere in an experimental diffraction tilt-pattern of bilayer graphene. Here, curves appear in pairs offset by a few degrees. This is most clearly seen in $\Gamma_2$ diffraction (Fig.~\ref{fig:blg}b \textit{pink, purple}) where the maximum intensity occurs at $\pm 1.1^\circ$. This angular distance in the split of paired tilt-patterns directly measures the Ewald sphere curvature and excitation error $s$:
$s = k_0-\sqrt{k_0^2-k_\rho^2} $
where $k_0$ is the wavenumber of the incident electron and $k_\rho$ is the in-plane radial distance to $k$-space origin. For small tilt angles and Bragg peaks close to the $k$-space origin this will scale approximately linearly, while for larger angles at larger radial distances a conversion from specimen tilt to a Cartesian basis is detailed in Supplemental Section III \cite{SuppRef}. Bragg rod intensity plots in $k_z$ corresponding to Figures~\ref{fig:blg}, \ref{fig:mos2}, and \ref{fig:mlg} are featured in Supplemental Figure S7 \cite{SuppRef}. Here, the $\pm 1.1^\circ$ split in the low-angle tilt-patterns corresponds to an excitation error of $0.085$ \invAA at 80 keV.

\section{Twisted, Moir\'e Layers}
\begin{figure}[ht]
    \includegraphics[width=1\linewidth]{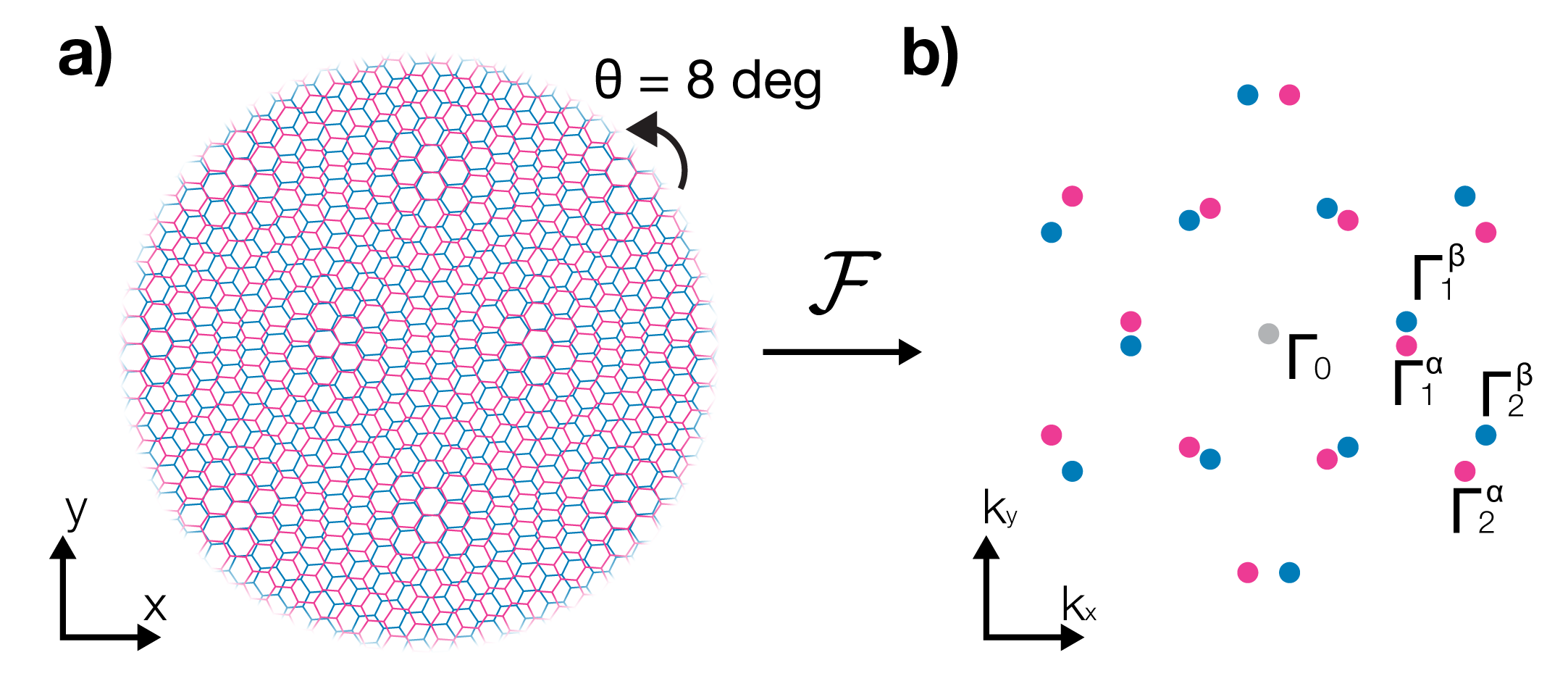}
    \caption{\textbf{Twisted Bilayer Graphene} a) Twisted bilayer graphene with an incommensurate intralayer twist angle ($\theta = 8$ deg). b) Reciprocal structure of incommensurate tBLG is a simple superposition of layers.}
    \label{FigtBLG}
\end{figure}

Significant interest in twisted multilayer materials has followed the micromechanical exfoliation of 2D heterojunctions \cite{fang2014strong} and discovery of superconductivity in low-twist angle bilayer graphene \cite{cao_unconventional_2018}. The reciprocal lattice of twisted bilayers is described by $|\mathcal{F}[\Sh_\alpha(\rbf)+\Sh_\beta(\rbf)]|^2 = |\Sh_\alpha(\kbf)|^2 + |\Sh_\beta(\kbf)|^2 + \Sh^*_\alpha(\kbf)\Sh_\beta(\kbf)+ \Sh_\alpha(\kbf)\Sh^*_\beta(\kbf)$, for layers $\alpha$ and $\beta$. For incommensurate stacking, the cross term is zero and the diffraction pattern is a trivial superposition of each individual layer (Fig.~\ref{FigtBLG}). This allows independent characterization of each incommensurate layer; however, we lose the ability to characterize interlayer spacing. If $\alpha$ and $\beta$ are commensurate \cite{MelePRB}, the cross term is zero where the Bragg rods from each layer do not overlap. Only overlapping rods may interfere and sinusoidally oscillate. As shown by Brown \textit{et al.}, each twisted layer can be independently mapped in real space with dark field TEM by placing an aperture around each distinct Bragg peak in the diffraction plane of the TEM \cite{brown_twinning_2012}. 

H. Yoo \textit{et al.} recently reported at low-twist angles ($<3$ deg) in bilayer graphene periodic restructuring occurs and superlattice peaks emerge \cite{yoo_atomic_2018}. Systems with periodic lattice distortions, either from interlayer interaction or charge order, are not so simply described as a superposition of layers\cite{hovden_atomic_2016}.

\begin{figure*}
    \includegraphics[width=1\linewidth]{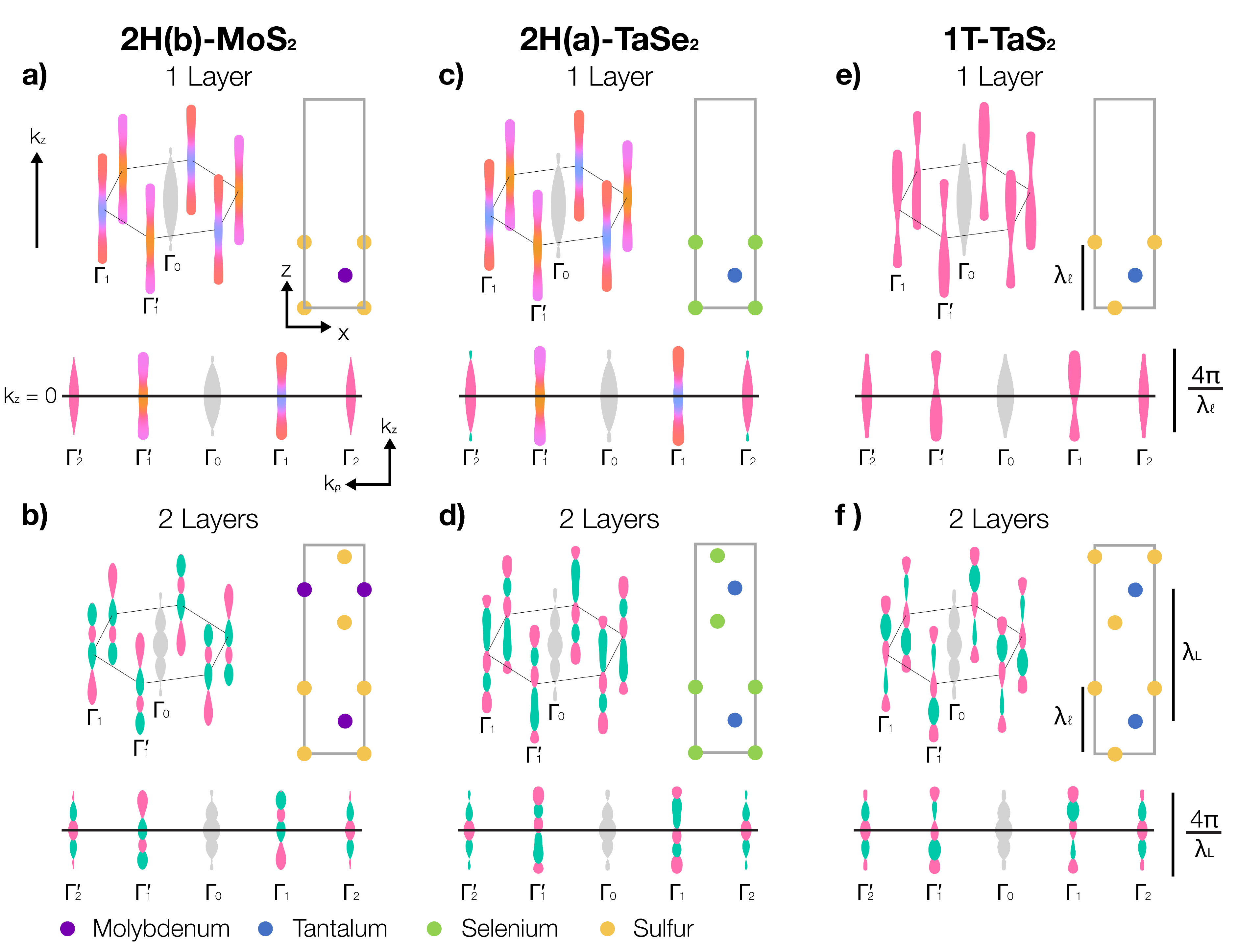}
    \caption{\textbf{3D reciprocal structure of 2D transition metal dichalcogenides and polytypes.} For each TMD, the Bragg rods ($\Gamma_0$, $\Gamma_1$) are shown in 3D alongside a real-space $<$100$>$ projection of the crystal stacking order. Below, a sideview of the Bragg rods ($\Gamma_0$, $\Gamma_1$, $\Gamma_2$) quantitatively illustrates the structure in $k$-space. Bragg rods have thickness and color indicating the complex magnitude and phase respectively. For single layer TMDs (a,c,e), two sinusoidal oscillations are determined by the interlayer spacing of the 3 atomic planes. The complexity increases noticeably for 2 vdW layers (b,d,f) which includes a beating frequency from interlayer spacing. Noticeably, H-phase MoS$_2$ and TaSe$_2$ have different stable multilayer stacking, denoted 2H(b) and 2H(a), leading to drastically different Bragg rod contours.}
    \label{fig:tmd}
\end{figure*}

\section{2D transition metal dichalcogenides}
\begin{figure}
    \centering
    \includegraphics[width=1\linewidth]{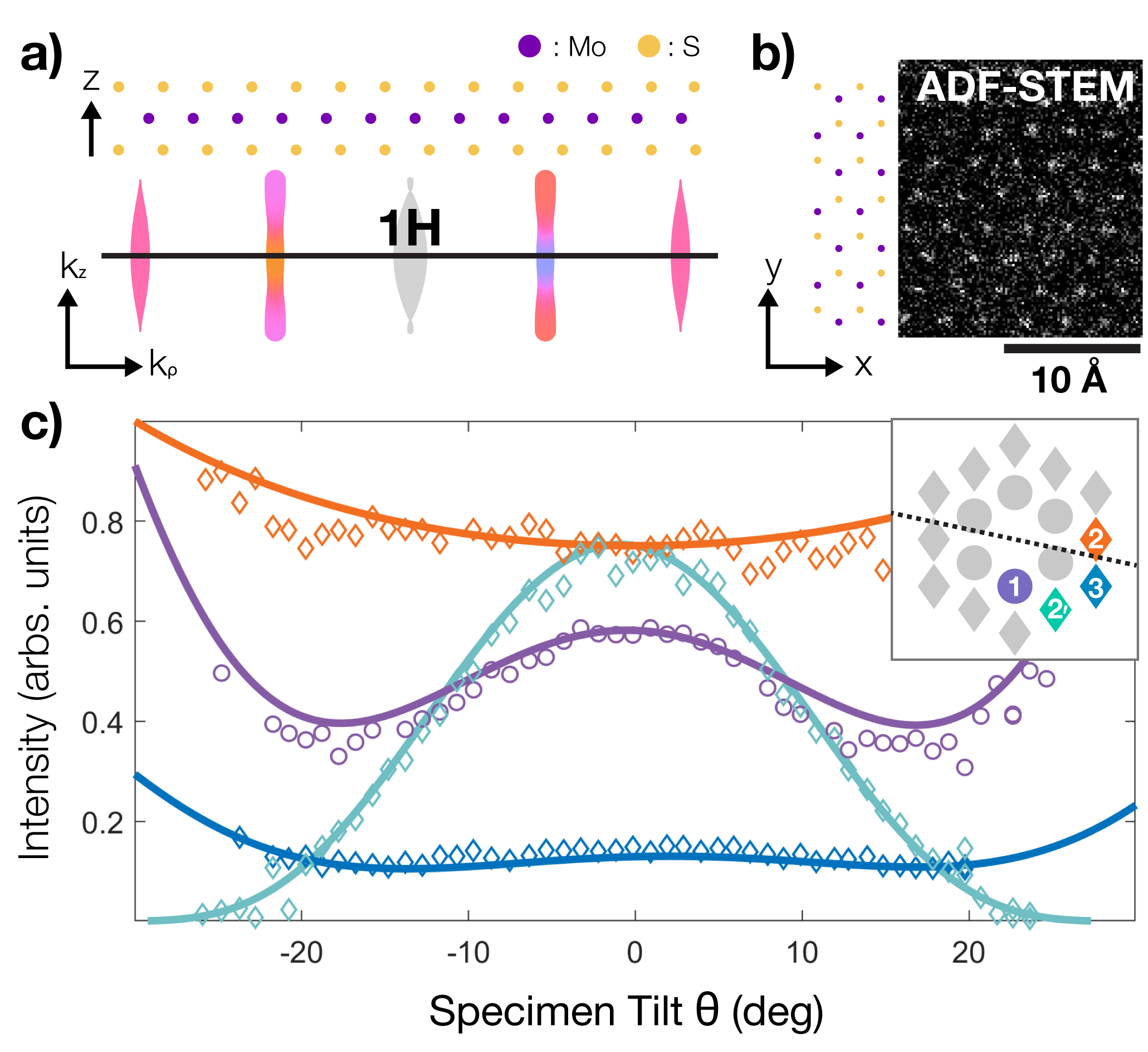}
    \caption{\textbf{K-structure of monolayer 2H-MoS$_2$.} The real and $k$-space structure of monolayer a) 2H-MoS$_2$ polytype shows mirror symmetry distinct from 1T. However, b) real-space schematic and ADF-STEM image along \zhat ~shows classification of 2H and 1T phase is extremely difficult because Mo atom intensities overwhelm S atoms. c) Directly probing the rod structure, the experimental tilt-pattern shows clear mirror symmetry and shows good agreement with the 2H analytic model. Rod intensity is plotted against $k_z$ in Supplemental Figure S7 \cite{SuppRef}.}
    \label{fig:mos2}
\end{figure}

Transition metal dichalcogenides (TMDs) are comprised of three atomic planes and two chemical species within each van der Waals (vdW) layer that add complexity to the Bragg rod structure (Fig. \ref{fig:tmd}a,c,e-\textit{top}). Six chalcogens encapsulate each metal atom geometrically with two tetrahedrons. Single layer TMDs are categorized into hexagonal `H' or trigonal `T' polytype phases by this local metal-chalcogen coordination complex~\cite{Wilson1969}. In the H-phase, the two tetrahedrons align along \zhat ~(Fig. \ref{fig:tmd}a), and in the T-phase, the two are displaced by 30 degrees giving rise to inversion symmetry ~(Fig. \ref{fig:tmd}e). Although isomeric to the 1T, the 2H phase notably breaks this inversion symmetry within a single layer but regains it in the bilayer. Broken inversion symmetry can significantly change electronic structure and has been associated with a metal-insulator transition in the 1T $\rightarrow$ 2H transformation \cite{acerce_metallic_2015, splendiani_emerging_2010} and the indirect to direct band gap transition in 2H TMDs reduced to a single layer (1H) \cite{splendiani_emerging_2010}. In several TMDs, such as TaS$_2$ and TaSe$_2$, the 1T phase permits room temperature charge ordering and even superconductivity at higher pressures \cite{sipos_mott_2008}.

Diffraction combined with specimen tilt can precisely determine metal-chalcogen coordination within a single vdW layer due to its sensitivity to crystal symmetry. 
The three atomic planes comprising a vdW layer are described by Bragg rods oscillating with a periodicity inversely proportional to $\lambda_{\ell}$, the intralayer spacing between chalcogen-chalcogen atomic planes:
\begin{align}
V_{1H}(\kbf) &= \Shb(\kbf)
[f_{m}(\kbf) + 2f_{c}(\kbf)e^{-i\kbf\cdot\rbf_0}\cos{(k_z \frac{\lambda_{\ell}}{2})}] \label{eq:1H}
\\
V_{1T}(\kbf) &= 
\Shb(\kbf)
[f_{m}(\kbf) + 2f_{c}(\kbf)\cos{(\kbf\cdot\rbf_0+k_z \frac{\lambda_{\ell}}{2})}]
\end{align}

where $f_{m}$ and $f_{c}$ are the atomic scattering factors of the metal and chalcogen atoms, respectively, and $\rbf_0$ is the in--plane metal--chalcogen bond direction ($\rbf_0 = \frac{\abf_1+\abf_2}{3}$). 1H denotes monolayer 2H.

Figure \ref{fig:tmd}-\textit{top} highlights the 3D reciprocal space structure of several monolayer TMDs. The change in metal-chalcogen coordination drastically changes the Bragg rod structure (Fig. \ref{fig:tmd} a,c vs. e), whereas the change in chemical composition alters the contour of the rod intensities (Fig. \ref{fig:tmd} a vs. c). The broken inversion symmetry of the 1H structure is represented in the complex phase of Eq.~\ref{eq:1H} that continuously changes on the $\Gamma_1$ Bragg rod along \kzhat ~(Fig.~\ref{fig:tmd} a,c)---this phase is not measurable from the diffraction amplitude. The 1T $\Gamma_1$ rods are markedly distinct with strong asymmetric oscillation of amplitude. Similar to graphene, we see TMDs possess $\Gamma_2$ rods symmetric about $k_z=0$ and insensitive to chalcogen coordination.

The experimental tilt-pattern of an exfoliated MoS$_2$ flake shown in Figure \ref{fig:mos2} reveals a single layer H phase. The $\Gamma_1$ and $\Gamma_3$ curves (Fig. \ref{fig:mos2}c-\textit{purple},\textit{blue}) are symmetric about $\theta = 0$ degree, which indicates a mirror plane at $k_z = 0$. This feature clearly discerns monolayers of the 2H and 1T polytypes (see also Supp. Fig. S3, S5) \cite{SuppRef}. The kinematic model of monolayer 2H-MoS$_2$ closely matches the experimental result (Fig. \ref{fig:mos2}c). Although monolayer 2H and 1T phases have different projected structure in real-space, the light sulfur atoms are virtually invisible in high-angle annular dark field (HAADF) STEM making this distinction challenging to characterize in real space (Fig. \ref{fig:mos2}b).

The intralayer spacing in a 2D TMD is precisely quantified by diffraction tilt-patterns for the first time. Nonlinear regression analysis of the experimental monolayer 2H-MoS$_2$ data reveals an intralayer chalcogen-chalcogen spacing ($\lambda_{\ell}$) of 3.07~\AA~with a 95\% confidence interval of $\pm$ 0.11~\AA~ based on a kinematic model. Multiple scattering may further reduce precision, especially in thicker systems containing strong scatterers. Our single layer value is comparable to the previously reported 3.01~\AA~ for bulk 2H-MoS$_2$ \cite{schonfeld_anisotropic_1983}.

The addition of a second vdW layer opens a wider range of stacking configurations and the Bragg rod complexity expands quickly---with 3 Fourier coefficients per vdW layer. Most notably, bilayer gains a beat frequency described by the interlayer spacing, $\lambda_{L}$. The interlayer beating is concisely expressed for bilayer 1T: $V_{2T}(\kbf) = V_{1T}(\kbf) \cdot 2 \cos{(k_z \frac{\lambda_{L}}{2}})$. The rapid rod oscillation from the larger vdW gap ($\lambda_{L}>\lambda_{\ell}$) beats with intralayer oscillations to create a non-uniform spacing between amplitude minima and maxima.

Additionally, multilayer TMDs have several stacking geometries both within and between their vdW layers. For instance, 2H-MoS$_2$ and 2H-TaSe$_2$ have distinct structures, typically denoted as 2H(b) and 2H(a) respectively (Fig.~\ref{fig:tmd}b,d).
The Bragg rod structure for single layer and bilayer T and H phases are shown in Figure~\ref{fig:tmd}. Supplemental Figure S3 provides an atlas of TMD stacking geometries and illustrates the distinct structures in $k$-space that allow unique identification and quantification \cite{SuppRef}.

\section{multilayer 2D materials} \label{sec:MLG}
\begin{figure}
    \centering
    \includegraphics[width=1\linewidth]{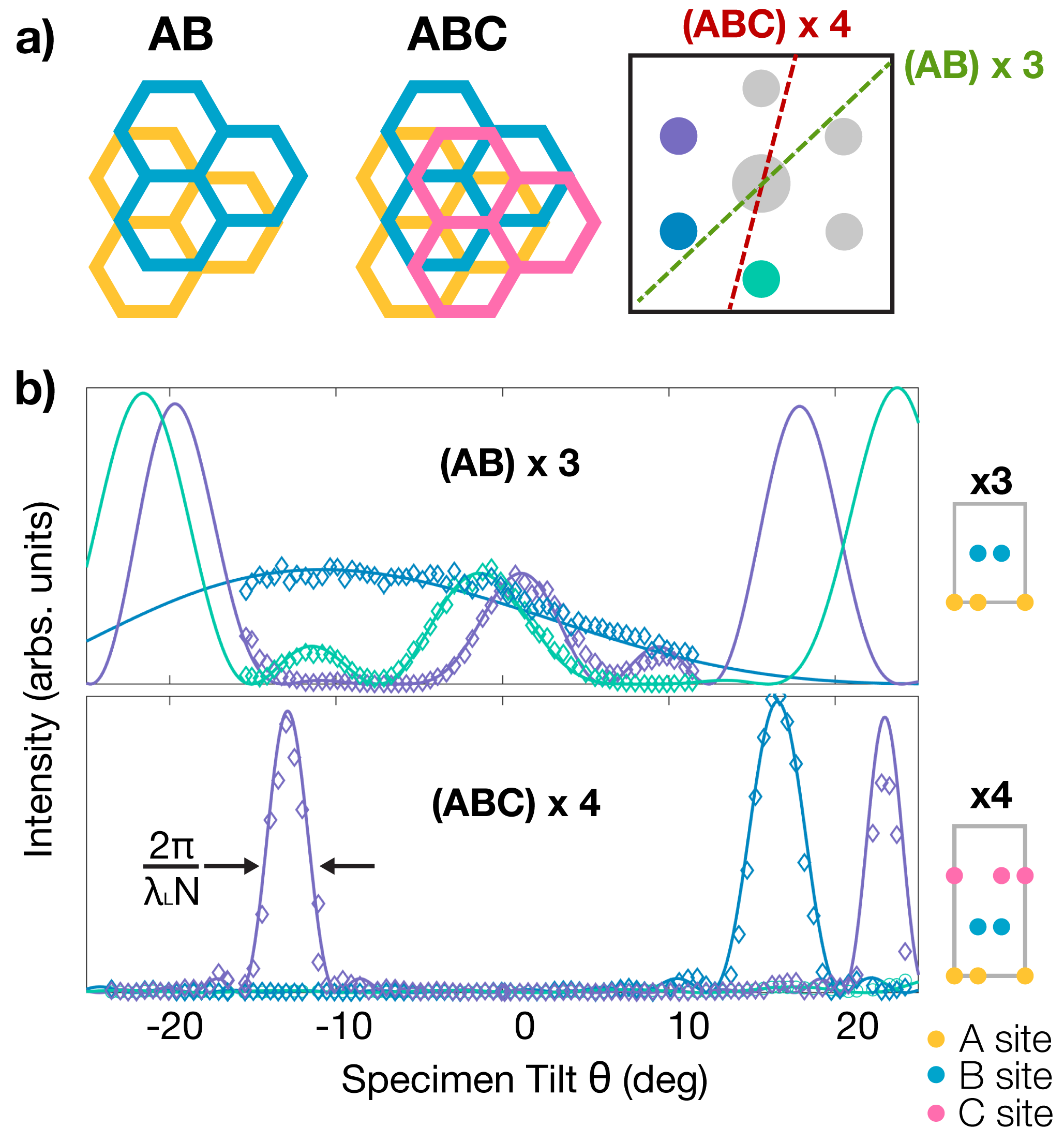}
    \caption{\textbf{Diffraction tilt-patterns of multilayer Bernal and rhombohedral graphene.} a) real-space stacking of Bernal (AB) and rhombohedral (ABC) graphene layers. b) Experimental diffraction tilt-patterns are plotted along with matched kinematically modeled patterns. Top right inset labels the plotted Bragg rods and specimen tilt axis. Rod intensity is plotted against $k_z$ in Supplemental Figure S7 \cite{SuppRef}.}
    \label{fig:mlg}
\end{figure}

Here we use multilayer graphene to illustrate how diffraction tilt-patterns can characterize thicker 2D materials. In atomically registered multilayer graphene, there are three possible sublattice positions---A,B,C---each one bond-length apart from the others (Fig. \ref{fig:mlg}a).
The two ordered stackings, hexagonally symmetric AB (Bernal) and rhombohedrally symmetric ABC, have been shown to have dramatically different band structures and transport properties \cite{xiao2011density,bao2011stacking}. However, thickness and stacking order determination is particularly difficult for samples more than three layers thick. In bulk materials, the rods give way to discrete peaks along \kzhat (Supp. Fig. S1), but at intermediate thicknesses (3--15 layers) they still contain interpretable out-of-plane structural information \cite{SuppRef}. Although the possible stacking configurations grows exponentially with thickness, leveraging minimal prior knowledge about the specimen significantly reduces the number of possibilities and makes exact determination of structure tractable.

Here, we characterize the out-of-plane structure of mechanically exfoliated 6-- and 12--layer graphene samples. At these intermediate thicknesses, the number of graphene layers is redundantly described by the width of each Bragg rod along \kzhat~($\Delta k_z = \frac{2\pi}{\lambda_L N}$), the angle which the 2nd order peak first reaches zero while tilting with an axis of rotation along $\Gamma_1$ ($N = \frac{21}{\theta(deg)}$), and the intensity of the second order Bragg Rod ($I=4\frac{\sin^2{\text{\textonehalf} k_z \lambda_L N}}{\sin^2{\text{\textonehalf}k_z \lambda_L}}$). These three relationships are derived in Supplemental Section IV from analytic models of multilayer graphene \cite{SuppRef}.

By measuring the relative intensity of the 1st and 2nd order Bragg peaks ($|\Gamma_1/\Gamma_2|$) at zero tilt ($k_z=0$), we can determine the fraction of each sublattice in the system. For instance, with equal number of all three sublattices' layers (e.g. ABCABC) the first order Bragg peaks have zero intensity; if the system has only two sublattices' layers in equal number (e.g. ABAB) the relative intensity is 0.25 (Supplemental Section VI) \cite{SuppRef}. 

Applying these rules to the tilt-pattern in Figure \ref{fig:mlg}b-\textit{top}, we determined the sample has 6 layers and an equal number of A and B sublattices. Registered 6-layer graphene has $3^5$ possible configurations. Eliminating the trivial duplicates and those with incorrect sublattice proportions leaves only 7 possible stacking orders from which we matched the correct stacking---ABABAB---by comparison with kinematically modeled tilt-patterns. 

Likewise, the sample in Figure \ref{fig:mlg}b-\textit{bottom} was found to be 12 layers thick with an equal proportion of A,B, and C sublattices, allowing the stacking order to be classified as ABCABCABCABC; the rhombohedral ordered stacking. Fast identification of rhombohedral graphene may have importance in fabrication of 2D heterostructure devices.

\section{summary \& conclusion}

Dimensionally confined 2D materials have rich 3D structure in reciprocal space described by near-infinite Bragg rods that oscillate with complex magnitude encoding the out-of-plane structure. Using a simple kinematic model of diffraction, combined with specimen tilt, the structure of these Bragg rods has been mapped in detail for several 2D materials (graphene, TMDs) across a range of stacking geometries. Using this 3D diffraction technique, we probed out-of-plane structure and symmetry to quantitatively determine critical structural parameters such as inter- \& intra-layer spacings and stacking order in multilayer graphene and TMDs. For single layer MoS$_2$ we extracted a chalcogen-chalcogen layer spacing of 3.07 $\pm$ 0.11~\AA. We accurately characterized the full interlayer stacking order of bilayer to multilayer graphene (demonstrated up to 12 layers), as well as identified multilayer rhombohedral graphene. The physical and electronic properties of layered 2D materials are often dramatically susceptible to these parameters. Although efficiently extracted with 3D diffraction, out-of-plane features are challenging or impossible to extract using real-space optical or surface measurement methods. However, our results are obtained using a rudimentary TEM available at most institutions. With the increasing complexity of multilayered materials, engineered by composition, twist, and stacking order---the foundational details outlined in this manuscript enable rapid and / or high-precision characterization across the complete class of 2D materials. Reciprocal structures illustrated throughout the manuscript and supplemental materials provide a 2D materials atlas for 3D diffraction. Furthermore, this work directly empowers a broader range of advanced diffraction based imaging techniques---such as dark-field TEM and 4D STEM---capable of mapping structural order in real space.

\begin{acknowledgments}
Authors thank David A. Muller for useful scientific discussion. S.H.S and N.S. contributed equally to this work. This work made use of the University of Michigan Center for Materials Characterization (MC2) as well as the Cornell Center for Materials Research (CCMR) an National Science Foundation MRSEC (DMR 1120296).
\end{acknowledgments}

\bibliography{refs}

\end{document}